\documentclass[aps, prl, twocolumn, showpacs
]{revtex4-1} 

\usepackage{amsmath}
\usepackage{amssymb}
\usepackage{graphicx}
\usepackage{color}

\renewcommand{\vec}[1]{\boldsymbol{#1}}
\begin{document}

\title{Extended propagation of powerful laser pulses in focusing Kerr media}
\author{V.~M.~Malkin}
\affiliation{Department of Astrophysical Sciences, Princeton University, Princeton, NJ USA 08540}
\author{N.~J.~Fisch}
\affiliation{Department of Astrophysical Sciences, Princeton University, Princeton, NJ USA 08540}

\date{\today}

\begin{abstract}
Powerful incoherent laser pulses can propagate in focusing Kerr media much
longer distances than can coherent pulses, due to the fast phase mixing that
prevents transverse filamentation.
This distance is limited by  4-wave scattering, which accumulates waves at
small transverse wavenumbers, where phase mixing is too slow to retain the
incoherence and thus prevent the filamentation.
However,  we identify how this theoretical limit can be overcome by countering
this accumulation through transverse heating of the pulse by random
fluctuations of the refractive index.
In these new regimes,  the laser pulse propagation distances are
significantly extended, making feasible a new class of random lasers, in
particular, ultra-powerful random lasers in plasmas.

\end{abstract}
\pacs{42.65.Sf, 42.65.Jx, 52.35.Mw}
\maketitle

Propagation of powerful laser pulses in focusing nonlinear media depends on the competition between the transverse dispersive spreading and focusing nonlinearity \cite{1964-PRL-Chiao,1964-IzVUZRadFiz-Talanov,1965-PRL-Kelley,1966-JETPLet-Bespalov,1968-UFN-Akhmanov-self-focusing,1993-PhysD-Malkin}. For negligible nonlinearity, the dispersion would double the cross section of a coherent pulse of  transverse size $L_\perp$ within the Rayleigh length $L_R\sim L_\perp^2/\lambda$, where $\lambda$ is the laser wavelength.
A strong enough nonlinearity would noticeably reduce the pulse cross section, or even cause transverse filamentation, within the length
$L_{SF}\sim v_g/\omega_{nl}$, where $\omega_{nl}$ is the nonlinear frequency shift  and  $v_g$ is the group velocity of the pulse. For $L_R<L_{SF}$, the dispersive spreading outruns the self-focusing and suppresses it. We will consider focusing Kerr-like media for which the nonlinear frequency shift is proportional to the pulse intensity  $I$,  namely, $\omega_{nl}=-\alpha I$ with the positive coefficient $\alpha > 0$. Then, $L_R/L_{SF}\sim P_{ch}/P_{cr}$,  where $P_{ch}\sim I L_\perp^2$ is the coherent pulse power and $P_{cr}\sim \lambda v_g/\alpha$ is the critical power of self-focusing.

A coherent laser pulse of power $P$ much greater than the critical power, $P\gg P_{cr}$, would experience transverse filamentation within a propagation length not much exceeding the self-focusing length $L_{SF}$. In contrast to this, incoherent laser pulses could traverse focusing Kerr-like media, remaining statistically uniform in the transverse directions, at arbitrarily large powers. What actually matters for non-filamentation of incoherent pulses is not the total power  $P$, but  the coherent sub-power $P_{ch}\sim I L_\perp^2$ located within the transverse correlation length $L_\perp$. If this sub-power is much smaller than the critical power, $P_{ch}\ll P_{cr}$, then the Rayleigh length is much shorter than the self-focusing length, so that nonlinear phase shift accumulated within the Rayleigh length is small $\phi_{nl}\sim \omega_{nl}L_R/v_g\sim P_{ch}/P_{cr}\ll 1$. 

The small parameter $P_{ch}/P_{cr}\ll 1$ enables a kind of perturbation theory.
Physically, this small parameter implies that the phase mixing of different waves occurs much faster than the nonlinear interaction. In zero-order approximation, the phases of different waves are random.
Statistical averaging over nearly random phases leads to a closed evolution equation for the pair correlation function of waves. This equation is usually referred to as a kinetic equation for waves. 
The pair correlation function of waves in Wigner representation is usually referred to as the wave spectral density. In general, the kinetic equation for waves contains terms linear, quadratic, cubic, and so forth in the wave spectral density. 

The linear term describes the  linear wave propagation. In conservative uniform media without group velocity dispersion, the linear propagation does not change the wave spectral density in the reference frame moving with the group velocity of the pulse. 

The wave spectral density variation rate, associated with the quadratic term in kinetic equation for waves, takes into account the statistically averaged nonlinear frequency shift, proportional to the pulse intensity. The respective nonlinear phase shift can bend phase fronts and cause self-focusing of incoherent pulses in a way similar to that of coherent pulses. The evolution of transverse size (radius) of an incoherent laser pulse within such a model was discussed in many papers (see, for instance, Refs. \cite{1974-JETP-Pasmanik, 1991-SovPhysUsp-Aleshkevich}).  
This approximation is sufficient for Kerr media in which the nonlinear frequency shift is proportional to the laser field intensity integrated over time or space, because such an integration effectively accomplishes statistical averaging (see, for instance, Ref. \cite{1999-PRL-Bang}).
However, for Kerr media of interest here in which the nonlinear frequency shift is proportional to the laser field intensity itself, higher order nonlinear processes can be important. 

In fact, the higher order nonlinear processes appear to be dominant for sufficiently powerful incoherent laser pulses statistically smooth in the transverse directions, namely, nearly uniform within the transverse correlation lengths of waves. This is because,
for statistically uniform pulses, the averaged nonlinear frequency shift does not vary in the transverse directions, so that there is no bending of the phase fronts. In other words, the quadratic term in the kinetic equation for waves is zero. The dominant nonlinear term in the kinetic equation for the waves is then the cubic term. 

The pulse power $P$, for which the cubic term is dominant, can be roughly evaluated as follows:
For pulses statistically varying in the transverse directions within the aperture size only (but not within much smaller sizes),
the wave spectral density variation rate, associated with the quadratic term in kinetic equation for waves, contains the small factor $P_{ch}/P$, while the variation rate, associated with the cubic term, contains the small factor $P^2_{ch}/P^2_{cr}$. The cubic process is much faster than the quadratic process for $P\gg P^2_{cr}/P_{ch}$. Of major interest here are pulses of very high powers $P$, perhaps exceeding the critical power $P_{cr}$ by a factor of a million, while, to secure the pulse incoherence, the ratio $P_{cr}/P_{ch} $  needs to be just somewhat larger than 1. 
The effect of the quadratic term in the kinetic equation for the waves on the propagation of such powerful pulses is then negligible.

The cubic term in kinetic equation for waves takes into account the 4-wave scattering. The evolution of wave spectral density due to the 4-wave scattering was subject to a long standing theoretical controversy, as recounted and resolved analytically in Ref. \cite{1996-PRL-Malkin}. That analytical resolution was recently supported numerically \cite{2015-PRE-Falkovich}. 

It is noteworthy that the 4-wave kinetic equation does not depend on the sign of the Kerr coefficient $\alpha$, which enters there only as $\alpha^2$. Being the same for the focusing and defocusing Kerr nonlinearities, the 4-wave kinetic equation cannot itself produce self-focusing and filamentation effects. Our initial spectrum satisfies the kinetic equation applicability condition with a large safety margin. There is, however, a possibility of the spectrum evolution that deteriorates the kinetic equation applicability and eventually ruins the phase randomness through a kind of Bose-Einstein condensation of waves at small transverse wavenumbers. 

Here we propose how to prevent an excessive wave accumulation at small transverse wavenumbers, thus extending the propagation of powerful laser pulses in focusing Kerr media beyond the known theoretical limits \cite{1996-PRL-Malkin}. We argue that the extension of pulse propagation length is indeed possible and may be very substantial in media with random fluctuations of the refractive index. Furthermore, we show how these substantially extended propagation lengths could make feasible a new class of random lasers \cite{2008-NPhys-random-lasers, 2010-NPhoton-Turitsyn, 2015-NComms-Churkin}, based on backward Raman amplification \cite{1966-PRL-Maier,1969-PhysRev-Maier,Malkin_99_PRL,Malkin_05_POP} and capable of reaching relativistic non-focused intensities  in plasmas.

As was already noticed above, the wave spectral density variation rate, associated with the cubic term in kinetic equation for waves, contains the small factor $P^2_{ch}/ P^2_{cr}$ compared to the rate of phase mixing.
This means that the propagation length, at which the wave spectral density starts to change noticeably due to the 4-wave scattering, is $P^2_{cr}/ P^2_{ch}\gg 1$ times larger than the Rayleigh length,
\begin{equation}\displaystyle
L_{4w}\sim L_{R} \, P_{cr}^2/P_{ch}^2\sim L_{SF}^2/L_{R} \, . 
\label{1}
\end{equation}
This length is  $L_{SF}/L_{R}\sim P_{cr}/ P_{ch}\gg 1$ longer than the length $L_{SF}$ within which a coherent pulse of the same intensity would break into filaments. 

A single 4-wave scattering of two waves with transverse wavenumbers $k_{\perp}\sim 2\pi/L_{\perp}$ into two new waves can produce a wave of about twice the  transverse wavenumber along with a wave of a small wavenumber. Further scattering produces waves of even larger transverse wavenumbers, so that the wave distribution spreads over a circle of growing radius $k_{\perp M}$ in the $\vec k_\perp$-plane. At the same time, the number of waves with small transverse wavenumbers increases. For smaller $k_\perp$, the Rayleigh length $L_R\propto L^2_{\perp}\propto k^{-2}_\perp$  is longer, while the 4-wave scattering length (\ref{1}) is respectively shorter.
Due to the larger rates at smaller $k_\perp$, the 4-wave scattering tends to establish the equilibrium Rayleigh-Jeans distribution of wave spectral density $N_{\vec k_\perp}\propto (k^2_\perp + k^2_{\perp m})^{-1}$ inside the circle $k_\perp\lesssim  k_{\perp M}$. This multi-scale distribution has roughly 
\begin{equation}\displaystyle
\Lambda={\rm ln }(k^2_{\perp M}/k^2_{\perp m})
\label{2}
\end{equation}
different scales each of which carries approximately the same fraction 
$1/\Lambda$ of total pulse intensity $I$.
Such a multi-scale optical turbulence does not satisfy the classical Kolmogorov hypothesis of spectral locality of interactions. There is, however, a more general kind of locality  \cite{1996-PRL-Malkin}, which modifies the estimate (\ref{1}) as follows:
\begin{equation}\displaystyle
L_{4wk_\perp}\sim L_{SF}^2\Lambda^2/(\Lambda_{k_\perp} L_{Rk_\perp}) \, ,
\label{3}
\end{equation}
where $\Lambda_{k_\perp}\approx {\rm ln }(k^2_\perp/k^2_{\perp m})$, for $k_\perp\gg k_{\perp m}$, and $\Lambda_{k_{\perp m}}\approx 1$. The modifying factor comes from the rate of 4-wave scattering.
Since this rate is quadratic in the wave intensity, two wave intensities contribute to it. One of the contributing waves must have wavenumber about $k_{\perp}$, and thus can carry the intensity $\sim I/\Lambda$, 
while the second wave may have any wavenumber smaller than $k_{\perp}$, and thus can carry the intensity $\sim I\Lambda_{k_\perp}/\Lambda$. Therefore, the rate of 4-wave scattering is modified by the factor $\sim \Lambda_{k_\perp}/\Lambda^2$, and the length of 4-wave scattering is modified by the inverse factor $\sim \Lambda^2/\Lambda_{k_\perp}$.

Apart from the ``number of waves", the 4-wave scattering conserves also the first and the second moments of the wave spectral density in $\vec k_\perp$-space. The first moment, having the physical meaning of the ``transverse momentum of waves", will be further assumed be zero. The second moment will be further referred to as the ``transverse energy of  waves", $I_\perp\propto \int N_{\vec k_\perp}k^2_\perp d^2{\vec k_\perp}$.  Though conserved in the  4-wave scattering, this quantity can grow due to the pulse scattering by random inhomogeneities of the medium. 
The transverse energy is located mostly at  $k_\perp\sim k_{\perp M}$, so that $I_\perp\propto  k^2_{\perp M}/\Lambda$. For an initial pulse parameters $I_{\perp 0}$, $k^2_{\perp 0}$ and $\Lambda_0\sim 1$, it follows $I_\perp\sim I_{\perp 0} k^2_{\perp M}/(\Lambda k^2_{\perp 0})$,  or
\begin{equation}\displaystyle
\Lambda\sim (I_{\perp 0}/{I_\perp}) k^2_{\perp M}/k^2_{\perp 0}\, .
\label{4}
\end{equation}
The length of $k^2_{\perp M}$ growth due to the 4-wave scattering is 
$L_{4M}\sim L_{4wk_{\perp M}}\sim L_{SF}^2\Lambda/L_{Rk_{\perp M}}\propto k^4_{\perp M}/I_\perp$.
Integration of the equation $d k^2_{\perp M}/dz \sim k^2_{\perp M}/L_{4 M}$ gives
\begin{equation}\displaystyle
k^4_{\perp M}/k^4_{\perp 0}\sim \int dz I_\perp /(I_{\perp 0}L_{40})\, ,
\label{5}
\end{equation}
where $L_{40}\sim L_{SF}^2/L_{Rk_{\perp 0}}$ is the initial length of 4-wave scattering.
The quantity $k_{\perp m}$ is then determined from formulas (\ref{2}) and (\ref{4}) for $\Lambda$. 

The applicability condition of random phase approximation, $L_{4k_\perp}> L_{Rk_\perp}$, is the most restrictive at $k_{\perp } \sim k_{\perp m}$, where it is equivalent to $L_{Rk_{\perp m}}< L_{SF}\Lambda$. This can be rewritten, using (\ref{2}) and (\ref{4}), in the form 
\begin{equation}\displaystyle
\exp\Lambda < ( L_{SF}/L_{Rk_{\perp 0}})\Lambda^2 I_{\perp }/{I_{\perp 0}} .
\label{6}
\end{equation}
If the transverse energy of waves were conserved, $I_{\perp}=I_{\perp 0}$, formulas (\ref{4}-\ref{5}) would give $\Lambda\sim k^2_{\perp M}/k^2_{\perp 0}\sim \sqrt{z/L_{40}}$. The applicability condition (\ref{6}) would restrict then the pulse propagation length as follows:  
\begin{equation}\displaystyle
z< z_*\sim L_{40}\, {\rm ln}^2 (P_{cr}/P_{ch 0})\, .
\label{7}
\end{equation}
This recovers the result of \cite{1996-PRL-Malkin}.
To remove this restriction, we propose to provide sufficient transverse heating of the pulse, which would enable $ \Lambda$ satisfying (\ref{6}) and nearly constant at $z>z_*$. The solution of equations (\ref{4}-\ref{5}) with nearly constant $ \Lambda$ is
\begin{equation}\displaystyle
z/L_{40}  \sim \Lambda^2 I_\perp/I_{\perp 0}\sim \Lambda k^2_{\perp M}/k^2_{\perp 0}\, .
\label{8}
\end{equation}
The applicability condition (\ref{6}) takes then the form 
\begin{equation}\displaystyle
\exp\Lambda < z/ L_{SF}\, .
\label{9}
\end{equation}

The largest transverse wavenumber $k_{\perp M}$, given by (\ref{8}), stays much smaller than the longitudinal wavenumber $k\sim 2\pi/\lambda$, so that the pulse remains paraxial, at
\begin{equation}\displaystyle 
z\ll \Lambda L_{40} k^2_{\perp 0}/k^2 \sim \Lambda L^2_{SF}/\lambda\, .
\label{10}
\end{equation}
This allows propagation distances $L_{SF}/\lambda\gg 1$ times larger than the filamentation length $L_{SF}$ of coherent pulses of the same intensity.

The transverse heating, needed to counter the tendency to Bose-Einstein condensation in 4-wave scattering, may be naturally produced by pulse scattering on random inhomogeneities of the refractive index. Inhomogeneities do not necessarily need to be statistically uniform.
A number of different arrangements might be possible. For example, sandwich structures could be used, wherein uniform layers of width, say, $ L_{40}$ alternate layers with random inhomogeneities of the refractive index on which the transverse heating occurs. In any event, random pushes in transverse wavenumbers for about $k_{\perp 0}/\Lambda$ within each propagation length about $ L_{40}$ would provide the transverse heating sufficient to preserve the pulse randomness and avoid the filamentation throughout distances much exceeding the previous theoretical limit (\ref{7}).

Consider, for example, statistically uniform random inhomogeneities of the refractive index having typical amplitude of relative fluctuations $f_r$, transverse correlation length $L_{\perp r}$ and longitudinal correlation length $L_{\parallel r}$. Let also, for simplicity, the group velocity of the laser pulse be about its phase velocity. Then, according to the Hamilton equations for rays, the transverse wavenumber typically changes, during the ray passing a single fluctuation, by 
$\delta k_\perp \sim k f_r L_{\parallel r}/L_{\perp r}$. This estimate assumes that the ray transverse travel distance, during the longitudinal travel distance $L_{\parallel r}$, does not exceed $L_{\perp r}$, which assumption is justified for $L_{\perp r}\gtrsim L_{\parallel r} k_\perp/k$. 
Within the propagation length $L_{40}\gg L_{\parallel r}$, a ray typically experiences  $L_{40}/ L_{\parallel r}$ random changes $\delta k_\perp$ of the transverse wavenumber. The resulting random change is typically $\delta k_\perp \sqrt{L_{40}/ L_{\parallel r}}\,$.  
In order for this to be about $k_{\perp 0}/\Lambda$, the system parameters should satisfy the condition
\begin{equation}\displaystyle 
\Lambda f_r (L_{SF}/L_{\perp r}) \sqrt{L_{\parallel r}/\lambda} \sim 1\, .
\label{11}
\end{equation}
 
Particularly interesting applications of these new regimes could be envisioned for powerful short laser pulses in plasmas where the Kerr effect is associated with the relativistic electron nonlinearity and laser intensities and powers could be extremely large.  
The dispersion law for electromagnetic waves in uniform plasmas is
$\displaystyle \omega^2=k^2c^2+\omega_e^2\, , $ 
where $c$ is the speed of light in vacuum and $\omega_e$ is the electron plasma frequency,
$\displaystyle \omega_e^2={4\pi e^2n_e }/{m_e}\, ,$ 
$-e$ is the charge of electron, $n_e$ is the electron concentration of plasma and $m_e$ is the mass of electron. 
For laser frequencies much larger than electron plasma frequency, $\omega\gg \omega_e$, the laser group velocity is close to the speed of light in vacuum, $v_g\approx c$, and the group velocity dispersion is small.

For a mildly relativistic electron motion in the laser field, the electron mass depends on the electron quiver velocity $v$ as 
$\displaystyle m_e={m}/{\sqrt{1-{v^2}/{c^2}}}\approx m \left(1+0.5{v^2}/{c^2}\right) .$  
The  normalized quiver energy, averaged over the laser period, $\overline{v^2}/c^2$ can be expressed in the terms of laser intensity $I$ as
$\displaystyle {\overline{v^2}}/{c^2}={4\pi e^2 I}/(m^2c^3\omega^2)\,  . $ 
The laser frequency shift due to the relativistic electron nonlinearity is then
$\displaystyle \omega_{nl}=-\alpha I\, , $ where
$\displaystyle \alpha ={\pi e^2 \omega_e^2}/(m^2c^3\omega^3)\, . $ 
For such a Kerr coefficient $\alpha$, the critical self-focusing power is 
$\displaystyle P_{cr}\sim (2m^2c^5/e^2)\,\omega^2/\omega_e^2 \approx 
17\times 10^{9} \, {\rm W }\, \omega^2/\omega_e^2$ . 
Though apparently only roughly estimated here, this $P_{cr}$, in fact, well agrees with the exactly calculated critical power for the relativistic self-focusing of axisymmetric laser pulses in plasmas \cite{Litvak1969,Max1974,Sun1987}.

For laser-to-plasma frequency ratio, say, 25, the critical power is $P_{cr}\sim 10^{13}$ W. To  have a weakly-nonlinear regime, take the power located within the transverse coherence size to be, say, 10 times smaller, $P_{ch0}\sim 0.1 P_{cr} \sim 10^{12}$ W. For the initial transverse wavenumber spread $ k_{\perp 0}\sim 0.03\, k$, the initial transverse correlation length is roughly $30$ laser wavelengths, $L_{\perp 0} \sim 30\, \lambda$. For $\lambda\sim 0.3\,\mu$m, this means $L_{\perp 0} \sim 10\,\mu$m. Then, the laser pulse intensity is $I\sim P_{ch0}/L_{\perp 0}^2\sim 10^{18} \, {\rm W/cm^2}$. A coherent pulse of such an intensity would experience transverse filamentation instability with the growth length $\displaystyle L_{SF}\sim (L_{\perp 0}^2/\lambda)\,  P_{cr}/P_{ch0} \sim 3\,$mm. The incoherent pulse propagates, not much changing the wave spectral density due to the 4-wave interaction, the distance $L_{40}\sim L_{SF}P_{cr}/P_{ch0} \sim 3\,$cm.  In a uniform plasma, at a few times larger distance, $ z_*\sim L_{40} \,{\rm ln^2}\left(P_{cr}/P_{ch0}\right)\sim 20\,$cm, the pulse incoherence would be ruined by the Bose-Einstein condensation leading to the pulse transverse filamentation. In a plasma with statistically uniform inhomogeneities of the refractive index, satisfying the condition (\ref{11}), the pulse can propagate, staying statistically uniform in the transverse directions and paraxial, a significantly larger distance  $L_{SF}^2/\lambda \sim 30\,$m. 

The plasma concentration in this example is $n\sim 2\times 10^{19} \,\rm cm^{-3}$. Such plasmas might be produced by ionization of dense aerosols \cite{2013-PRL-Aerosols, 2014-JAerosolSci-fisch}. The droplet size can be as small as a few microns.
The amplitude of relative fluctuations of the refractive index  $f_r$ can be expressed in the terms of the amplitude of relative fluctuations of the plasma electron concentration, $f_n=\delta n/n$, as follows: $f_r=0.5f_n\omega_e^2/\omega^2$. 
For $L_{\parallel r}\sim L_{\perp r} \sim 10\,\mu$, $\Lambda\sim 2$ and other parameters the same as in the above example, the formula (\ref{11}) gives $f_n\sim 0.3$. For 10 times larger $L_{\parallel r}$ and the previous  values of all other parameters, the formula (\ref{11}) would give $f_n\sim 0.1$. 

These new extended propagation regimes are particularly important for backward Raman amplification of laser pulses in plasmas \cite{Malkin_99_PRL,Malkin_00_POP, Malkin_14-EPJST}, for which the transverse filamentation instability, so far, was one of major limiting factors. An appropriate randomization can suppress harmful filamentation instability without noticeably affecting the useful amplification resonance.

However, even more important than improving upon the current resonant method of backward Raman amplification, the randomization makes feasible entirely different, random laser amplification regimes. 
This is critical for a number of reasons.

First, a practical reason is that it is technologically challenging to produce highly homogeneous plasmas wherein the useful amplification resonance is not much spoiled by the plasma density inhomogeneities. In plasmas with large random density inhomogeneities, the amplification occurs in the parts of the plasma where the resonance conditions are randomly met. 
This reduces the average amplification rate, thus increasing the amplification length. The new propagation regimes suggested here do allow such extended amplification lengths which were not achievable previously.  

Furthermore, the extended amplification lengths imply longer pump pulses, so that less intense pumps can carry even larger fluences. Lower pump intensities alleviate the limitation imposed by the effect of Langmuir wave breaking, devastating in deep wave-breaking regimes  \cite{Malkin_99_PRL,Malkin_14-EPJST,Toroker_14-POP,2015-POP-Edwards}. The pump laser pulse intensity at the threshold of  Langmuir wave breaking is $I_{br}=\pi m^2c^5/(16e^2\lambda^2)\,\omega_e^3/\omega^3 \approx 
1.7\times 10^{9} \, ({\rm W }/\lambda^2)\, \omega_e^3/\omega^3$. For the parameters from the above numerical example, this gives  $I_{br}\sim 10^{14}\, \rm W/cm^2$. For a pump of $I_{br}/2$ intensity, it would take about 60 cm, to amplify the counter-propagating seed pulse to the intensity  $10^{18}\, \rm W/cm^2$, while contracting to the duration 100 fs, at the energy transfer efficiency 50$\%$.

Furthermore, random amplification regimes are capable of overcoming the previous theoretical limit on the output pulse intensities, imposed by the relativistic electron nonlinearity that detunes and thus saturates the amplification \cite{Malkin_14-POP, Malkin_14-PRE}. The nonlinear detuning, even though it is varying along the pulse, can be randomly compensated by fluctuations in plasma frequency.

The input pulse randomization might be produced by scattering in a plasma layer with random density inhomogeneities, or through randomization techniques of the type developed for smoothing speckles in longer and less intense laser pulses used in direct irradiation of inertially confined targets for nuclear fusion \cite{1983-OpticCom-Lehmberg,1984-PRL-Kato,1989-JournApplPhys-Skupsky}.

The drawback of Raman amplification in the random laser regime is that the output pulse is not highly focusable.  However, this focusability can be regained by employing the two-step backward Raman amplification scheme \cite{Malkin_05_POP}.
 In such a scheme, the first step aims to accomplish the greatest possible longitudinal compression of the largest possible pump pulse energy. It requires large laser-to-plasma frequency ratio, because the achievable longitudinal contraction is an increasing function of this parameter. The randomization may be needed, only at this step, to suppress the transverse filamentation instability of the amplified pulse (which could otherwise reduce the allowed amplification length and pump energy longitudinal compression) and to enjoy the benefits mentioned above. The first-step output pulse is used then as a pump for the second step of the backward Raman amplification. Since the pump fluctuations are averaged when it is consumed by the counter-propagating amplified pulse, the requirements to the pump quality are very lenient. The second step aims primarily to regain the pulse focusability, while also accomplishing additional longitudinal compression of the laser energy, achievable in denser plasmas where the plasma frequency is larger and the plasma period is shorter. The amplification length at this step is relatively short, and the transverse filamentation instability of the amplified pulse does not pose a serious problem. Therefore, the randomization is not needed, and a highly coherent and focusable seed pulse can be used, preserving these properties through the amplification.

In summary, the key findings here are:  

1. New regimes of powerful laser pulse propagation in focusing Kerr media are found.  These regimes allow propagation distances significantly exceeding the previous theoretical limit.

2. These regimes can be used to improve the resonant backward Raman amplification of ultra-powerful laser pulses in plasmas with random density fluctuations at levels not much affecting the useful resonance, while suppressing the transverse filamentation instability.

3. These regimes make feasible a new class of random lasers, wherein the output pulses reach relativistic intensities at unprecedented large powers in plasmas with random density fluctuations at levels strongly affecting the useful amplification resonance.

This work was supported by DTRA HDTRA1-11-1-0037, by NSF PHY-1202162, and by
the NNSA SSAA Program under Grant No~ DE274-FG52-08NA28553.


\begin{thebibliography}{10}
\newcommand{\enquote}[1]{``#1''}
\expandafter\ifx\csname url\endcsname\relax
  \def\url#1{{#1}}\fi
\expandafter\ifx\csname urlprefix\endcsname\relax\def\urlprefix{}\fi
\vspace{-0.75cm}
\bibitem{1964-PRL-Chiao}
R.~Y. Chiao, E.~Garmire, and C.~H. Townes, \enquote{Self-Trapping of Optical
  Beams,} Phys. Rev. Lett. {\bf 13}, 479--482 (1964).
\newline http://dx.doi.org/10.1103/PhysRevLett.13.479

\bibitem{1964-IzVUZRadFiz-Talanov}
V.~I. Talanov, \enquote{On self-focusing of electromagnetic waves in nonlinear
  mediums,} Izv. VUZov. Radiophysica {\bf 7}, 564 (1964).

\bibitem{1965-PRL-Kelley}
P.~L. Kelley, \enquote{Self-Focusing of Optical Beams,} Phys. Rev. Lett. {\bf
  15}, 1005--1008 (1965).
\newline http://dx.doi.org/10.1103/PhysRevLett.15.1005

\bibitem{1966-JETPLet-Bespalov}
V.~I. Bespalov and V.~I. Talanov, \enquote{{Filamentary Structure of Light
  Beams in Nonlinear Liquids},} {JETP Letters} {\bf {3}}, {307} ({1966}).
\newline $\rm http://www.jetpletters.ac.ru/ps/1621/article\_24803.shtml$

\bibitem{1968-UFN-Akhmanov-self-focusing}
S.~A. Akhmanov, A.~P. Sukhorukov, and R.~V. Khokhlov, \enquote{Self focusing
  and diffraction of light in a nonlinear medium,} Soviet Physics Uspekhi {\bf
  10}, 609 (1968).
\newline http://dx.doi.org/10.1070/PU1968v010n05ABEH005849

\bibitem{1993-PhysD-Malkin}
V.~M. Malkin, \enquote{On the analytical theory for stationary self-focusing of
  radiation,} Physica D: Nonlinear Phenomena {\bf 64}, 251--266 (1993).
\newline http://dx.doi.org/10.1016/0167-2789(93)90258-3

\bibitem{1974-JETP-Pasmanik}
G.~A. Pasmanik, \enquote{Self-interaction of incoherent light beams,} 
Zh. Eksp. Teor. Fiz. {\bf 66}, 490 (1974) 
[Sov. Phys. JETP {\bf 39}, 234 (1974)].

\bibitem{1991-SovPhysUsp-Aleshkevich}
V.~A. Aleshkevich, G.~D. Kozhoridze, and A.~N. Matveev, \enquote{Self-action of
  partly coherent laser radiation,} Soviet Physics Uspekhi {\bf 34}, 777--803
  (1991).
\newline http://stacks.iop.org/0038-5670/34/i=9/a=R02

\bibitem{1999-PRL-Bang}
O. Bang, D. Edmundson, and W. Kr\'olikowski, \enquote{Collapse of Incoherent Light Beams in Inertial Bulk Kerr Media,} Phys. Rev. Lett. {\bf 83}, 5479--5482
  (1999).
\newline http://link.aps.org/doi/10.1103/PhysRevLett.83.5479

\bibitem{1996-PRL-Malkin}
V.~M. Malkin, \enquote{Kolmogorov and Nonstationary Spectra of Optical
  Turbulence,} Phys. Rev. Lett. {\bf 76}, 4524 (1996).
\newline http://dx.doi.org/10.1103/PhysRevLett.76.4524

\bibitem{2015-PRE-Falkovich}
G.~Falkovich and N.~Vladimirova, \enquote{Cascades in nonlocal turbulence,}
  Phys. Rev. E {\bf 91}, 041\,201 (2015).
\newline http://link.aps.org/doi/10.1103/PhysRevE.91.041201

\bibitem{2008-NPhys-random-lasers}
D.~Wiersma, \enquote{The physics and applications of random lasers,} Nature
  Physics {\bf 4}, 359--367 (2008).
\newline http://dx.doi.org/10.1038/nphys971

\bibitem{2010-NPhoton-Turitsyn}
S.~Turitsyn, S.~Babin, A.~El-Taher, P.~Harper, D.~Churkin, S.~Kablukov,
  J.~Ania-CastÃ£Ã³n, V.~Karalekas, and E.~Podivilov, \enquote{Random
  distributed feedback fibre laser,} Nature Photonics {\bf 4}, 231--235 (2010).
\newline http://dx.doi.org/10.1038/nphoton.2010.4

\bibitem{2015-NComms-Churkin}
D.~Churkin, I.~Kolokolov, E.~Podivilov, I.~Vatnik, M.~Nikulin, S.~Vergeles,
  I.~Terekhov, V.~Lebedev, G.~Falkovich, S.~Babin, and S.~Turitsyn,
  \enquote{Wave kinetics of random fibre lasers,} Nature Communications {\bf
  2}, 6214 (2015).
\newline http://dx.doi.org/10.1038/ncomms7214

\bibitem{1966-PRL-Maier}
M.~Maier, W.~Kaiser, and J.~A. Giordmaine, \enquote{Intense Light Bursts in the
  Stimulated Raman Effect,} Phys. Rev. Lett. {\bf 17}, 1275 (1966).
\newline http://link.aps.org/doi/10.1103/PhysRevLett.17.1275

\bibitem{1969-PhysRev-Maier}
M.~Maier, W.~Kaiser, and J.~A. Giordmaine, \enquote{Backward Stimulated Raman
  Scattering,} Phys. Rev. {\bf 177}, 580 (1969).
\newline 
http://link.aps.org/doi/10.1103/PhysRev.177.580

\bibitem{Malkin_99_PRL}
V.~M. Malkin, G.~Shvets, and N.~J. Fisch, \enquote{Fast compression of laser
  beams to highly overcritical powers,} Phys. Rev. Lett. {\bf 82}, 4448 (1999).
\newline http://dx.doi.org/10.1103/PhysRevLett.82.4448

\bibitem{Malkin_05_POP}
V.~M. Malkin and N.~J. Fisch, \enquote{Manipulating ultra-intense laser pulses
  in plasmas,} Phys. Plasmas {\bf 12}, 044\,507 (2005).
\newline http://dx.doi.org/10.1063/1.1881533

\bibitem{Litvak1969}
A.~G. Litvak, \enquote{Finite-amplitude wave beams in a magnetoactive plasma,}
  Zh. Eksp. Teor. Fiz. {\bf 57}, 629 (1969)  [Sov. Phys. JETP {\bf 30}, 344 (1970)].

\bibitem{Max1974}
C.~Max, J.~Arons, and A.~B. Langdon, \enquote{Self-modulation and self-focusing
  of electromagnetic waves in plasmas,} Phys. Rev. Lett {\bf 33}, 209 (1974).
\newline http://dx.doi.org/10.1103/PhysRevLett.33.209

\bibitem{Sun1987}
G.-Z. Sun, E.~Ott, Y.~C. Lee, and P.~Guzdar, \enquote{Self-focusing of short
  intense pulses in plasmas,} Phys. Fluids {\bf 30}, 526 (1987).
\newline http://dx.doi.org/10.1063/1.866349

\bibitem{2013-PRL-Aerosols}
M.~J. Hay, E.~J. Valeo, and N.~J. Fisch, \enquote{Geometrical Optics of Dense
  Aerosols: Forming Dense Plasma Slabs,} Phys. Rev. Lett. {\bf 111}, 188\,301
  (2013).
\newline http://link.aps.org/doi/10.1103/PhysRevLett.111.188301

\bibitem{2014-JAerosolSci-fisch}
D.~Ruiz, L.~Gunderson, M.~Hay, E.~Merino, E.~Valeo, S.~Zweben, and N.~Fisch,
  \enquote{Aerodynamic focusing of high-density aerosols,} Journal of Aerosol
  Science {\bf 76}, 115 -- 125 (2014).
\newline http://dx.doi.org/10.1016/j.jaerosci.2014.05.010

\bibitem{Malkin_00_POP}
V.~M. Malkin, G.~Shvets, and N.~J. Fisch, \enquote{Ultra-powerful compact
  amplifiers for short laser pulses,} Phys. Plasmas {\bf 7}, 2232 (2000).
\newline http://dx.doi.org/10.1063/1.874051

\bibitem{Malkin_14-EPJST}
V.~M. Malkin and N.~J. Fisch, \enquote{Key plasma parameters for resonant
  backward Raman amplification in plasma,} Eur. Phys. J. Special Topics {\bf
  223}, 1157 (2014).
\newline http://dx.doi.org/10.1140/epjst/e2014-02168-0

\bibitem{Toroker_14-POP}
Z.~Toroker, V.~M. Malkin, and N.~J. Fisch, \enquote{Backward Raman
  amplification in the Langmuir wavebreaking regime,} Physics of Plasmas {\bf
  21}, 113\,110 (2014).
\newline http://dx.doi.org/10.1063/1.4902362

\bibitem{2015-POP-Edwards}
M.~Edwards, Z.~Toroker, J.~Mikhailova, and N.~Fisch, \enquote{The efficiency of
  Raman amplification in the wavebreaking regime,} Physics of Plasmas {\bf 22},
  074\,501 (2015).
\newline http://dx.doi.org/10.1063/1.4926514

\bibitem{Malkin_14-POP}
V.~M. Malkin, Z.~Toroker, and N.~J. Fisch, \enquote{Saturation of the leading
  spike growth in backward Raman amplifiers,} Phys. Plasmas {\bf 21}, 093\,112
  (2014).
\newline http://dx.doi.org/10.1063/1.4896347

\bibitem{Malkin_14-PRE}
V.~M. Malkin, Z.~Toroker, and N.~J. Fisch, \enquote{Exceeding the leading spike
  intensity and fluence limits in backward Raman amplifiers,} Phys. Rev. E {\bf
  90}, 063\,110 (2014).
\newline http://dx.doi.org/10.1103/PhysRevE.90.063110

\bibitem{1983-OpticCom-Lehmberg}
R.~H. Lehmberg and S.~P. Obenschain, \enquote{Use of induced spatial
  incoherence for uniform illumination of laser fusion targets,} Optics
  Communications {\bf 46}, 27 -- 31 (1983).
\newline http://dx.doi.org/10.1016/0030-4018(83)90024-X

\bibitem{1984-PRL-Kato}
Y.~Kato, K.~Mima, N.~Miyanaga, S.~Arinaga, Y.~Kitagawa, M.~Nakatsuka, and
  C.~Yamanaka, \enquote{Random Phasing of High-Power Lasers for Uniform Target
  Acceleration and Plasma-Instability Suppression,} Phys. Rev. Lett. {\bf 53},
  1057--1060 (1984).
\newline http://dx.doi.org/10.1103/PhysRevLett.53.1057

\bibitem{1989-JournApplPhys-Skupsky}
S.~Skupsky, R.~W. Short, T.~Kessler, R.~S. Craxton, S.~Letzring, and J.~M.
  Soures, \enquote{Improved laser-beam uniformity using the angular dispersion
  of frequency-modulated light,} Journal of Applied Physics {\bf 66},
  3456--3462 (1989).
http://dx.doi.org/10.1063/1.344101

\end{thebibliography}

\providecommand{\noopsort}[1]{}\providecommand{\singleletter}[1]{#1}%

\end{document}